# Different Channels to Transmit Information in a Scattering Medium

Xuyu Zhang[1,2], Jingjing Gao[1,3], Yu Gan[1,3], Chunyuan Song[1,3], Dawei Zhang[2,*], Songlin Zhuang[2], Shensheng Han[1,3,4], Puxiang Lai[5,6,7,*], and Honglin Liu[1,3,6,*]

[1]*Key Laboratory for Quantum Optics, Shanghai Institute of Optics and Fine Mechanics, Chinese Academy of Sciences, Shanghai 201800, China*
[2]*Department of Physics, University of Shanghai for Science and Technology, Shanghai 200093, China*
[3]*Center of Materials Science and Optoelectronics Engineering, University of Chinese Academy of Sciences, Beijing 100049, China*
[4]*Hangzhou Institute for Advanced study, University of Chinese Academy of Sciences, Hangzhou 310024, China.*
[5]*Department of Biomedical Engineering, The Hong Kong Polytechnic University, Hong Kong SAR, China*
[6]*Hong Kong Polytechnic University Shenzhen Research Institute, Shenzhen 518000, China*
[7]*Photonics Research Institute, The Hong Kong Polytechnic University, Hong Kong SAR, China*

*[*]dwzhang@usst.edu.cn, puxiang.lai@polyu.edu.hk, and hlliu4@hotmail.com*

**Abstract**

A channel should be built to transmit information from one place to another. Imaging is 2 or higher dimensional information communication. Conventionally, an imaging channel comprises a lens and free spaces of its both sides. The transfer function of each part is known; thus, the response of a conventional imaging channel is known as well. Replacing the lens with a scattering layer, the image can still be extracted from the detection plane. That is to say, the scattering medium reconstructs the channel for imaging. Aided by deep learning, we find that different from the lens there are different channels in a scattering medium, i.e., the same scattering medium can construct different channels to match different manners of source encoding. Moreover, we found that without a valid channel the convolution law for a shift-invariant system, i.e., the output is the convolution of its point spread function (PSF) and the input object, is broken, and information cannot be transmitted onto the detection plane. In other words, valid channels are essential to transmit image information through even a shift-invariant system.

**Introduction**

   Communication is the basis of daily life and modern civilization. A channel is the path to transfer information from one place to another. How to transmit information through a channel in an optimal way, where, optimal means that the obtained code will determine the event unambiguously, isolating it from all others in the set, and will have minimal length, that is, it will consist of a minimal number of symbols, is well studied in information theory. It also provides methodologies to separate real information from noise and to determine the channel capacity required for optimal transmission conditioned on the transmission rate. However, all information has to been transferred into temporal dimension symbols first in information theory. Imaging is 2 or higher dimensional information communication. To enhance imaging capabilities, understanding of the imaging channel is also the key. Conventionally, an imaging channel comprises a lens and free spaces of its both sides. The transfer function of each part is known; thus, the response of a conventional imaging channel is known as well. Replacing the lens with a scattering layer, the image can still be extracted from the detection plane. That is to say, the scattering medium reconstructs the channel for imaging. What are the characteristics of the imaging channel in a scattering medium? Are there any differences from a conventional one?

   Deep learning has been widely applied in imaging through scattering media due to its data driven and physical-model free feature [1-7]. Usually, a trained neural network is only applicable to a particular scenario with limited generalization capability. Instead of a pure image extraction method we have demonstrated that deep learning can also be used as a tool to explore unknown principles in physics. With sufficient samples for training, it can drain out information encoded in random speckles [8]. Here, by designing an experiment and utilizing deep learning as a standard tool able to extract all encoded information with sufficient data, we found that a scattering medium could construct different channels to transmit image information, and its micro structure contributed a new type of channel. To further verify the discovery, it was hypothesized that image information could not be transmitted through a medium without such micro structure. We demonstrated the hypothesis with phase grids. Moreover, it is a basic law that an output of a shift-invariant system is the convolution of its point spread function (PSF) and an input target [9-11]. However, we found that without valid channel the convolution law was broken, the output did not equal the convolution of the PSF and the target, and image information could not be transmitted through the shift-invariant system.

In the following content, we first introduce the experiment and simulation setups in the method section. Next, the results and corresponding analyses are presented in the result section. Finally, we give some discussions and reach the conclusions.

**Methods**

The experimental setup is shown in Fig.1. In Fig.1(a), a green laser (MGL-III-532-200mW, Changchun New Industries Optoelectronics Tech) was expanded to illuminate a reflective digital micromirror device (DMD, V-7001 VIS, ViALUX) with a pixel pitch of 13.7 μm. Handwritten digits from the MNIST database [12-16] are used as the objects, which are reshaped into 64×64 arrays and loaded on the central 64×64 pixels of the DMD in subsequence. The reflected light propagated through an aperture with a diameter of 10 mm to a digital camera (DCU224M, Thorlabs). Instead of a lens, the aperture was used to construct the channel for imaging, and diffraction patterns were recorded on the camera. Here, the aperture and the free spaces of its both sides built up the channel. Next, we inserted a homemade 220-grit ground glass diffuser into the optical path just next to the aperture, and recorded the speckles patterns. The diffuser and the free spaces of its both sides built up the channel, the effective diameter of the diffuser is limited by the aperture. Since the free spaces are the same, for simplification, when we mention channel below, it only refers to the aperture or diffuser. In contrast, we inserted a rotating diffuser into the laser beam to create a pseudo thermal source, and repeated the data acquisition step for the aperture and diffuser, respectively. The obtained 4 groups of data, each has 20,000 pairs of objects and intensity patterns, 18000 patterns were used to train and 2000 to test a UNet neural network [17-19], respectively. From Ref.8 we know that after training with sufficient data the network can drain out image information encoded in a detected pattern. If no image is extracted from a pattern, it means no valid information in the recorded pattern. The object distance from DMD to the diffuser $z_2 = 16\ cm$, and the image distance from the diffuser to the camera $z_3 = 25\ cm$.

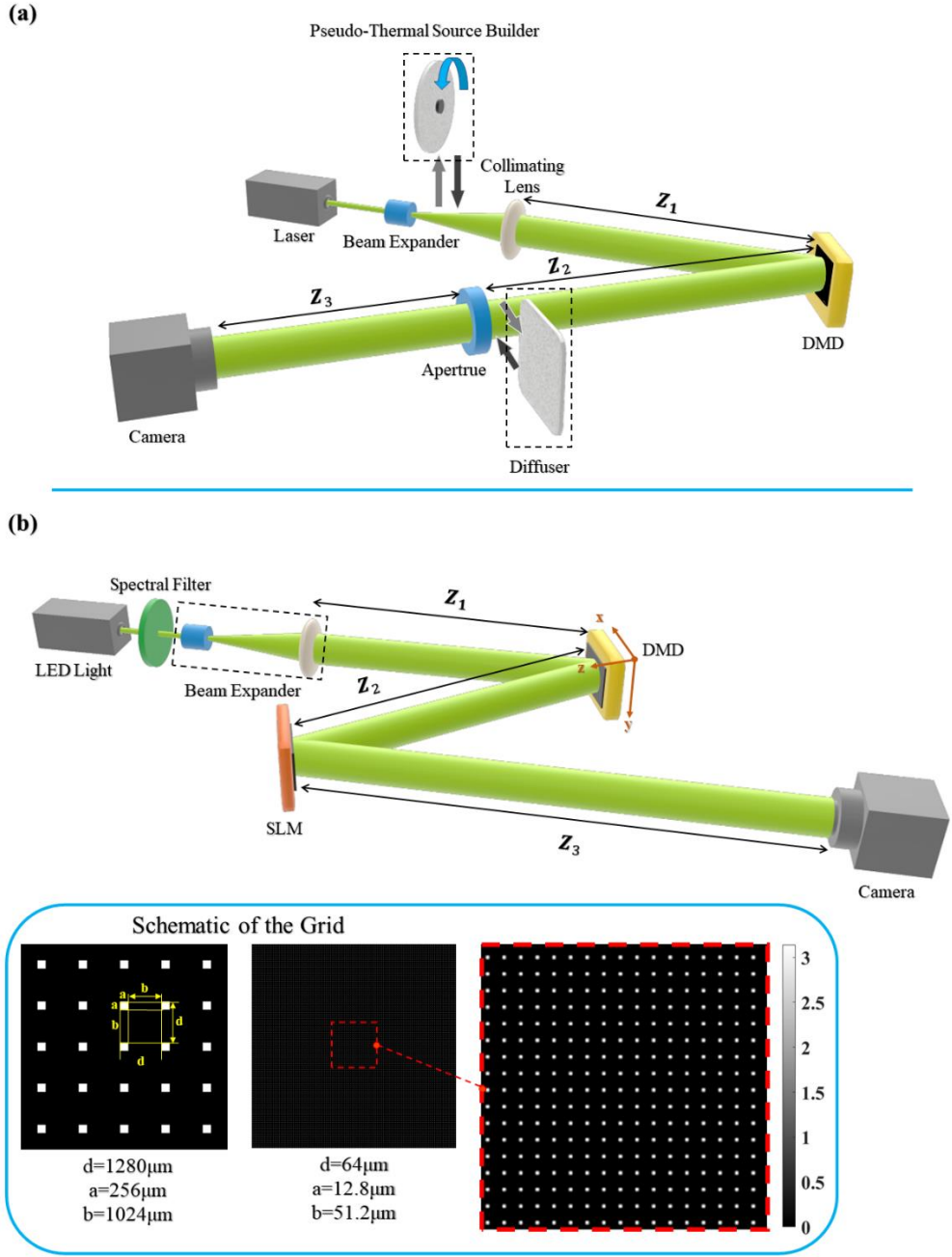

Fig. 1. Schematics of the experimental setups. (a) Experimental setup to test different channels in a scattering medium. The switching of channels is realized by inserting a rotating diffuser and a stationary diffuser in and out of the beam. (b) Experimental setup to demonstrate no valid channel in phase grids, although they constitute shift-invariant systems. DMD - digital micromirror device. SLM - spatial light modulator. The lower half shows the schematic maps of two phase-grids loaded on the SLM, respectively. $d$ is the cell period, $a$ is the width of the phase square with a value of $\pi$ in each cell. The insert on the right is the zoom in of the dash red square.

In the grid experiment as shown in Fig.1(b), in order to eliminate the rain effect in the recorded patterns caused by the rotating diffuser, a LED light (M530L4-C1, Thorlabs) coupled with a spectral filter (FL532-10, Thorlabs), whose central wavelength and linewidth are 532 nm and 10 nm respectively, was used as the illumination source. Different phase grids were displayed on a spatial light modulator (SLM, LETO-2, Holoeye, pixel size 6.4 μm), whose array size was 1920×1080. The schematics of the grids are shown in the down side of Fig.1(b). Two phase grids were used to replace the diffuser. One had the parameters of d=64μm, a=12.8μm, and b=51.2μm (Grid I), while the other had d=1280μm, a=256μm, and b=1024μm (Grid II). Due to the limited size of the SLM, the diameter of the grid occupied 1000 pixels, corresponding to 6.8 mm, slightly smaller than the size of the aperture in Fig.1(a). The reflected light from the DMD impinges on the SLM, then reaches the camera. The PSFs of the system and intensity patterns corresponding to different objects are recorded. The recorded intensity patterns are also used to train the UNet network [8], but no image could be predicted successfully. For comparison, we also loaded random phase map sharing the same statistical distribution of the diffuser onto the SLM, repeated the processes of the grid experiment.

The original recorded patterns are cropped into 1024×1024 arrays in following figures, and furtherly down sampled into 512×512 arrays for network training and testing.

For a spatial shift-invariant system, if the input function is $f(x)$ and the response of the system is $S\{\}$, we can get the output function $g(x)$ with a displacement $x_0$ as [20-22]

$$g(x - x_0) = S\{f(x - x_0)\}. \tag{1}$$

Then, for an optical imaging system with spatial shift-invariance, the output

$$I(x) = O(x) * PSF(x) = \int_{-\infty}^{\infty} O(x')PSF(x - x')dx', \tag{2}$$

where $O(x)$ is an intensity object and $PSF(x)$ denotes the PSF of the system. In other words, under incoherent illumination, the intensity pattern on the camera plane equals the convolution of the object and the PSF. In order to justify whether they are equal, their structural similarity (SSIM) was calculated [23,24].

$$SSIM(x, y) = \frac{(2\mu_x\mu_y + c_1)(2\sigma_{xy} + c_2)}{(\mu_x^2 + \mu_y^2 + c_1)(\sigma_x^2 + \sigma_y^2 + c_2)}, \tag{3}$$

where $\mu_x$ and $\mu_y$ are the averages, $\sigma_x^2$ and $\sigma_y^2$ are the variances of $x$ and $y$, respectively, $\sigma_{xy}$ is the covariance of $x$ and $y$. $c_1 = (k_1 L)^2$ and $c_1 = (k_1 L)^2$ are the constants used to maintain stability and avoid division by zero [25], and $L = 2^{Bit} - 1$ is the dynamic range of the pixel value. Generally, for 8 Bit data, the $L$ value is 255. By searching for information, generally it is most suitable to compare pictures when $k_1$ is set as 0.01 and $k_2$ as 0.03.

The same UNet as used in Ref.8 was adopted for image extraction. The inputs of the network are preprocessed 512×512 recorded patterns, not necessarily speckle patterns. The outputs of the trained network are the reconstructed images with an array size of 512×512. The UNet network uses Python compiler based on Keras/Tensorflow 2.0 Library, and the GPU edition of the training UNet is NVIDIA RTX 3060 laptop. The total number of training epochs is 50, and the learning rate at the beginning is set as 2×10$^{-4}$. After 5 epochs, if the loss value doesn't decrease, the learning rate will be adjusted to one tenth of the previous one until the learning rate is reduced to 2×10$^{-6}$. When the loss value does not decrease after 15 epochs, the training will be terminated.

Simulation based on wave optics is also applied to verify the hypothesis, considering its noise-free advantage. The configuration and parameters are the same to the experiment. In simulation, we created 500 frames of independent speckle illuminations on the object to achieve incoherent illumination.

## Results

Under coherent illumination, the trained UNet can reconstruct the image well (see Fig.2), the aperture is a valid channel to transmit image information from the object plane to the camera plane. After inserting the diffuser, although the recorded patterns are seemly random speckles, images can still be predicted well by the network, which means that image information is still delivered to the detection plane. That is to say, the channel functions properly. The main difference is that the refractive index distribution of the diffuser encodes the aperture channel, as also inferred from Ref.8. Under incoherent illumination, through the aperture channel, no image can be extracted by deep learning. Apparently, there is no image information in detected patterns, and the aperture channel fails in transmitting information. In this case the aperture is an invalid channel. However, after inserting the diffuser in, images can be extracted from recorded intensity patterns, which means that a new type of channel appears due to the diffuser to enable successful information transmission. We believe that the new type of channel is formed by the micro structure of the diffuser. To avoid the effect of projection, we must bear in mind that the size of the target should not be large.

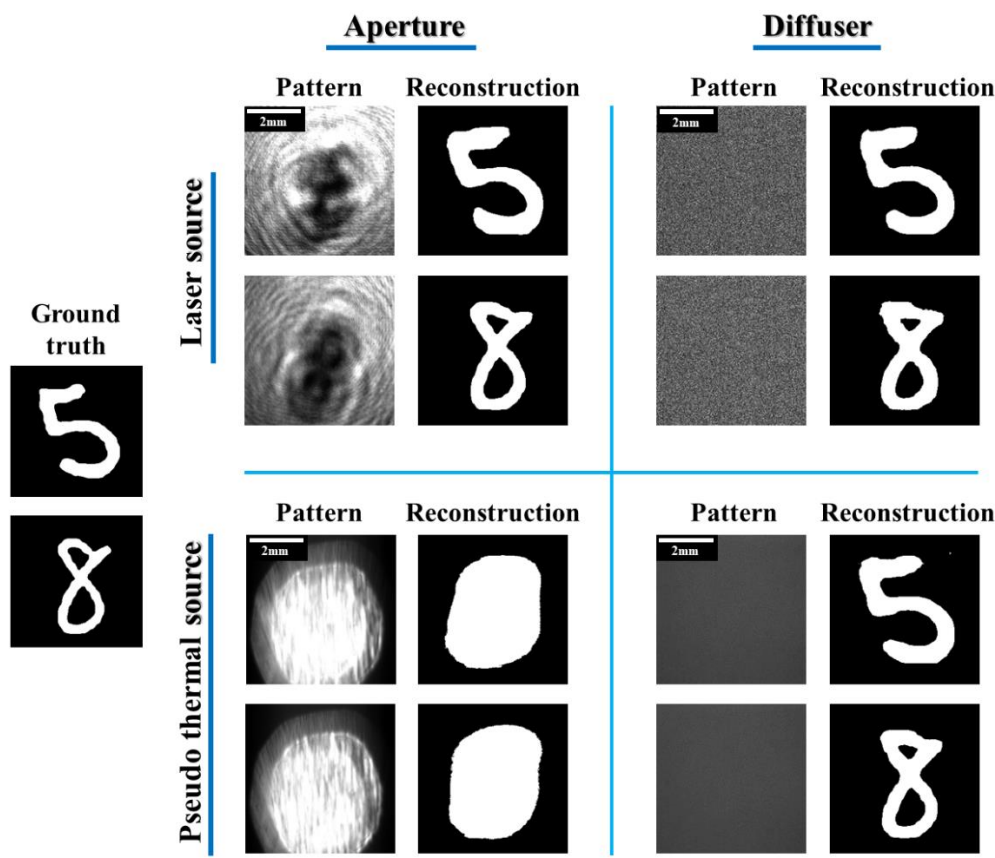

Fig. 2 Experimental results of different channels to transmit information in a scattering medium. Two digits "5" and "8" are used as ground truths. For each case the recorded patterns and reconstructed images are shown. Under coherent illumination of the laser, the recorded patterns after the aperture are the diffraction patterns of input objects. The trained network can certainly reconstruct the corresponding images. While after the diffuser, the camera records high contrast speckle patterns, the aperture channel is encoded by the randomly distributed refractive index on the diffuser. Information can still be delivered to the detection plane and extracted by deep learning. Under incoherent illumination of the pseudo thermal source, no image can be extracted from the pattern after the aperture. In contrast, images can be well predicted from the low contrast speckle patterns after the diffuser. The rain effect on the recorded patterns in the aperture case is due to the rotating of the diffuser to generate the pseudo thermal source.

The experimental results for the further demonstration with phase grids are shown in Fig.3. Row A in Fig.3(a) shows the recorded PSFs when a point was displayed on the DMD. As the point shifts, the PSF remains its shape but shifts accordingly. The grid system is shift-invariant, confirmed in Fig.3(b) as well.

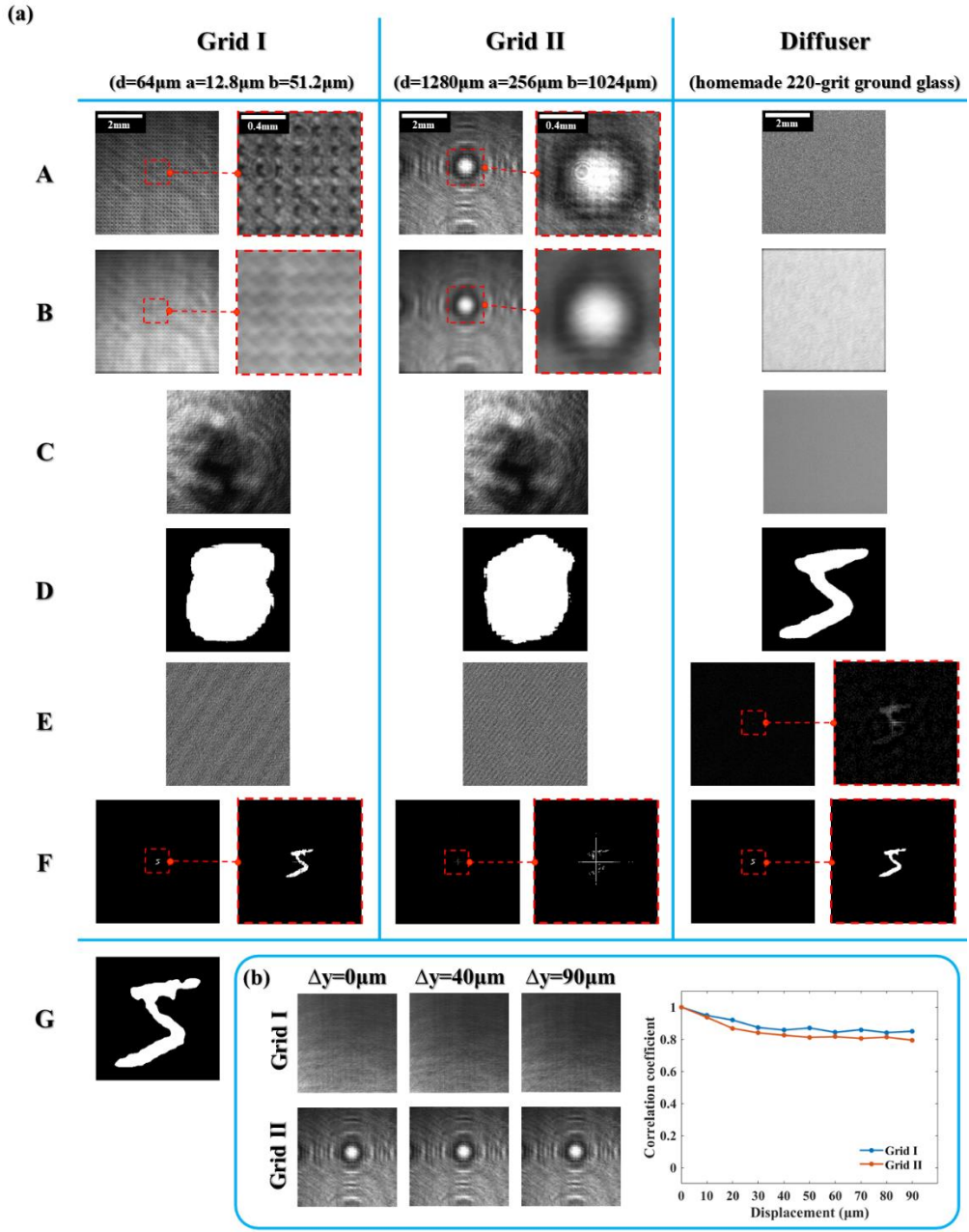

Fig.3 Experimental results of grid and diffuser channels under incoherent illumination. (a) Row A shows the PSFs of different systems. From the inserted local zoom in we can see many detail structures. Row B is the convolution of the digit 5 and the PSF correspondingly. The recorded patterns are shown in Row C. Rows D, E and F are the extracted images by deep learning from recorded patterns in Row C, by deconvolution from Rows C and B, respectively. Deep learning shows great advantage in extracting image from a speckle pattern with noise, the reconstructed image in Row D is much better than in Row E. The inserts show the zoom in of each dash square. The image of the input object is shown in Row G. (b) Recorded PSFs of grids 1&2 at different point positions, which confirm that the grid systems are shift-invariant. The correlation coefficient curves are flat over the scan range, which is already bigger than input objects.

According to the convolution law, the output on the camera plane for an object on the DMD should be the convolution of the object and the PSF. But the recorded patterns in Row C are different from the calculated convolution patterns in Row B in Fig.3(a). Comparing to the diffuser, the SSIM factors of recorded patterns and the convolutions for the grid are significantly lower (see Table 1). After deconvolution we can obtain the image from the convolution patterns but the recorded patterns. And the trained UNet network cannot reconstruct the image from the recorded pattern, either. While the recorded patten is similar to the convolution pattern for the diffuser. Based on these evidences, we can say that the phase grid doesn't have the micro structure of the diffuser, provides no valid channel to transmit information, and there is no image information in the recorded pattern. Moreover, without valid channel, the convolution law for the shift-invariant system fails. We know that images can be extracted by deep learning even when objects go beyond the memory effect ranges [26-29], i.e., the shift-invariance doesn't hold any more [30-37]. Apparently, channel is the essential route to transmit information.

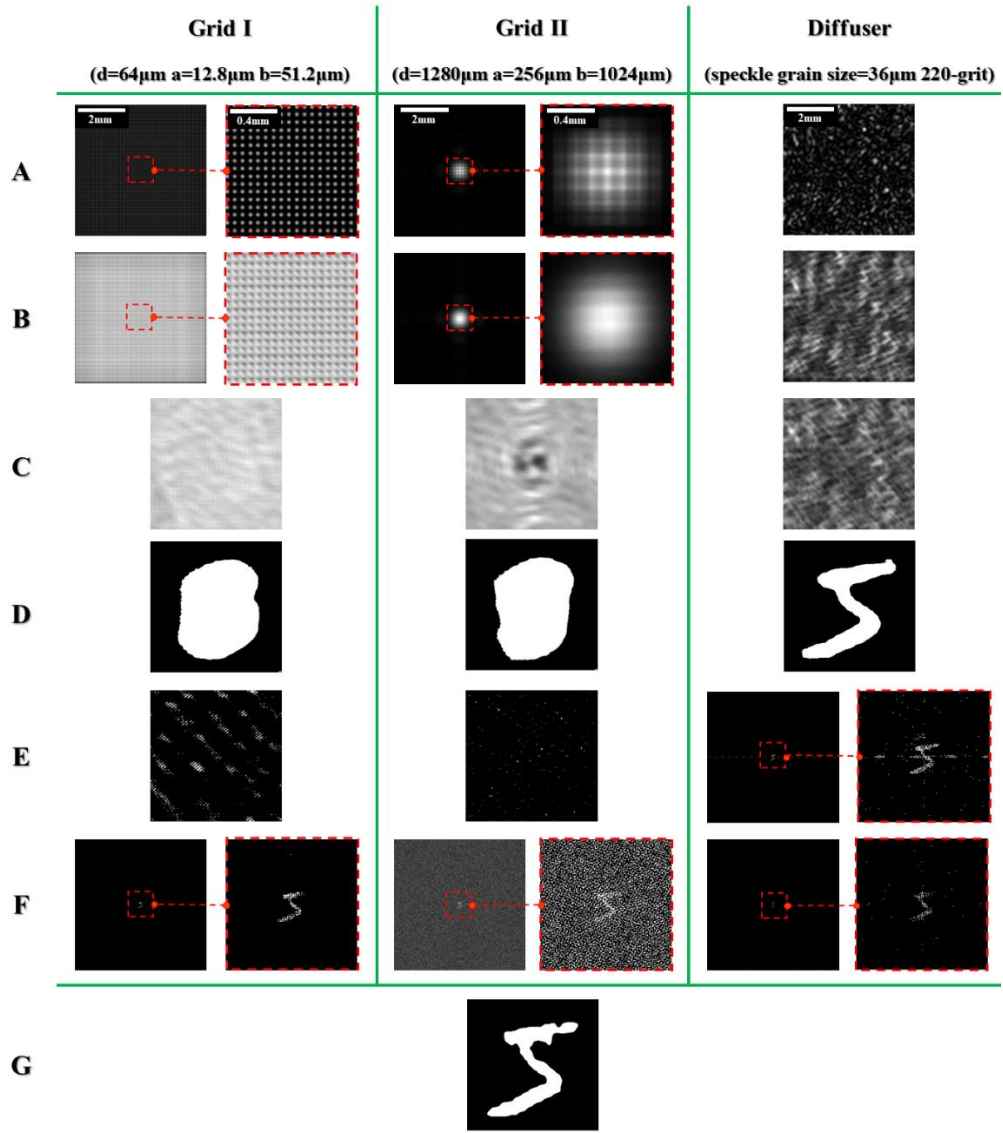

Fig.4 Corresponding simulation results of grid and diffuser channels under incoherent illumination. No image can be reconstructed from recorded patterns by either deep learning or deconvolution for both grids. There is no valid channel in the grid to transmit information, although the system constructed by it is shift-invariant.

Fig.4 shows the simulation results, in comparison to the experimental results in Fig.3. Due to less noise, the patterns in both Rows B and C have better contrast to see their fine structures. For the diffuser, the two patterns look quite similar, while for the grids, such similarity disappears.

Table 1 SSIM between the recorded pattern and calculated convolution pattern

|  | Grid I | Grid II | Diffuser |
|---|---|---|---|
| Experiment | 0.2024 | 0.2456 | 0.7534 |
| Simulation | 0.2092 | 0.0285 | 0.6804 |

In Table 1, the SSIM values of the two patterns for the diffuser are obviously bigger than the two grids. The reason for the extremely low SSIM=0.0285 for grid II is that in simulation the convolution pattern has higher intensity in and around the center and a dark background while it is reversed for the intensity distribution in the recorded pattern. The diffuser provides valid channel to transmit information, so the recorded pattern can be considered as equal to the convolution of the input object and the PSF. The unique micro structure of the diffuser, rather than the structure of the grids, build the channel to transmit image information under incoherent illumination.

**Discussions and Conclusions**

One may argue that the wide spread of the PSF is the reason of the failure in image reconstruction for the aperture in Fig.2, i.e., the resolution on the recorded plane is not sufficient to differentiate the structure of the image. If it is correct, we should be able to reconstruct images from the recorded patterns in both Figs.3&4, since the fine structures of the PSFs of the grids provide the resolution to distinguish small details, as shown in the deconvolution results. In fact, we cannot extract any information about the object from the recorded pattern by either deep learning or deconvolution. Hence, the failure of imaging is due to lack of effective information in the recorded pattern but not the limited resolution.

In information theory, there are different ways for source coding. In imaging, coherent and incoherent illuminations provide two different ways to encode object information. In coherent illumination, the information is encoded on wavefront, while in incoherent illumination, it is encoded on intensity. In order to effectively transmit image information, the channel should match the coding way of the source. The micro structure of the diffuser constructs a new type of channel efficient to transmit information encoded by incoherent illumination.

Channel bandwidth is critical to determine the quality of reconstructed image. Scattering media, like apertures, lenses and other optical components, construct imaging channels, and different channels in the medium has different capacities. Understanding imaging from the information angle may bring breakthroughs to some long pursued but yet unsolved problems. The discovery of different channels is the first step to unveil the mask on information transmission in imaging. Next, we shall build different channel models, study channel capacities, and establish an information theory for imaging.

In summary, deep learning is applied to discover unknown channels in scattering media, one more evidence that it is a powerful tool to study unknown physical principles and/or mechanisms. With the help of deep learning, we find that channel is essential to transmit image information, and there are new types of channels built by its micro structure in a scattering medium. Without valid channel, light can transmit through the media but not the information, even for shift-invariant systems. The results revolutionize our understanding of both deep learning and imaging, and may have profound impacts in many areas.

**Funding.** National Natural Science Foundation of China (NSFC) (81930048, 81671726, 81627805); Hong Kong Research Grant Council (25204416); Hong Kong Innovation and Technology Commission (ITS/022/18, GHP/043/19SZ, GHP/044/19GD); Guangdong Science and Technology Commission (2019BT02X105 and 2019A1515011374).

**Acknowledgements.** H. L. conceived the idea and designed the experiment and simulation. X. Z., J. G., Y. G., and C. S. implemented the experiment and simulation. X. Z., P. L. and H. L. analyzed the data. All contributed in writing and revising the manuscript.

**Competing Interests.** The authors declare no conflict of interests.